\documentclass[a4paper,10pt]{article}
\pdfoutput=1 

\usepackage{jcappub} 
 \usepackage{indentfirst}
\usepackage{appendix}

\title{Can the $H_{0}$ tension be resolved in extensions to $\Lambda$CDM cosmology?}

\author[a]{Rui-Yun Guo,}
\author[a]{Jing-Fei Zhang,}
\author[a,b,c,1]{Xin Zhang\note{Corresponding author.}}
\affiliation[a]{Department of Physics, College of Sciences, Northeastern University, \\Shenyang
110819, China}
\affiliation[b]{Ministry of Education Key Laboratory of Data Analytics and Optimization \\for Smart Industry, 
Northeastern University, Shenyang
110819, China}
\affiliation[c]{Center for High Energy Physics, Peking University, \\Beijing 100080, China}

\emailAdd{guoruiyun110@163.com, jfzhang@mail.neu.edu.cn, zhangxin@mail.neu.edu.cn}

\abstract{We wish to investigate whether there is an extension to the base $\Lambda$CDM cosmology that can resolve the tension between the Planck observation of the cosmic microwave background anisotropies and the local measurement of the Hubble constant. We consider various plausible extended models in this work, and we use the Planck 2015 observations, combined with the baryon acoustic oscillation data, the JLA type Ia supernovae data, and the local measurement of the Hubble constant (by Riess et al. in 2016), to make an analysis. We find that the holographic dark energy plus sterile neutrino model can reduce the tension to be at the 1.11$\sigma$ level, but this model is obviously not favored by the current observations. Among these extended models, the $\Lambda$CDM+$N_{\rm eff}$ model is most favored by the current observations, and this model can reduce the tension to be at the 1.87$\sigma$ level. By a careful test, we conclude that none of these extended models can convincingly resolve the $H_0$ tension.}

\begin{document}
\maketitle
\flushbottom

\section{Introduction}
\label{sec1}

In the past few decades, by accumulating large amount of accurate measurement data of distance--redshift relation and large-scale structure of the universe, a prototype of the standard cosmological model, i.e., the $\Lambda$ cold dark matter ($\Lambda$CDM) model, has been established. In the $\Lambda$CDM model, dark energy is provided by a cosmological constant $\Lambda$ (equivalent to the vacuum energy density) and dark matter is cold. It has been found that a spatially-flat $\Lambda$CDM cosmology with purely adiabatic, Gaussian initial fluctuations can explain and fit various observational data quite well. In particular, the observation of the Planck satellite mission~\cite{Ade:2015xua} strongly favors a basic 6-parameter $\Lambda$CDM cosmology.

However, in recent years, it was found that some cracks appear in the $\Lambda$CDM cosmology in the aspect of observation. For example, using the Planck observation of the cosmic microwave background (CMB) power spectra, the base $\Lambda$CDM model predicts a lower value of the Hubble constant, compared to the local measurement based on the method of distance ladder.
In 2016, Riess et al.~\cite{Riess:2016jrr} gave a result of the local measurement of the Hubble constant, $H_{0}=73.00\pm1.75$ km s$^{-1}$ Mpc$^{-1}$ (hereafter R16), which is $3.3\sigma$ higher than the fitting result of $66.93\pm0.62$ km s$^{-1}$ Mpc$^{-1}$ predicted by the Planck collaboration~\cite{Aghanim:2016yuo} assuming the $\Lambda$CDM model with 3 neutrino flavors having two massless neutrinos and a mass of 0.06 eV. The $H_{0}$ tension between the R16 result and the Planck data has attracted lots of attention of cosmologists. On one hand, the distance ladder measurement has reduced the uncertainty (of R16) to 2.4\%, which is a significant improvement compared to the previous local measurements of $H_{0}$ with the 3--5\% uncertainty~\cite{Riess:2011yx,Sorce:2012pk,Freedman:2012ny,Suyu:2012aa}. On the other hand, the $\Lambda$CDM fitting result of $H_0$ given by the Planck observation has a less than 1\% precision~\cite{Ade:2015xua}. Thus, both the two methods give precision measurements, but they are in significant, more than 3$\sigma$, tension.

The $H_0$ tension has stimulated some serious investigations on the possible systematic errors in either the Planck observation or the local measurement, but all these efforts failed to identify any obvious problem with either analyses~\cite{Spergel:2013rxa,Addison:2015wyg,Aghanim:2016sns,Efstathiou:2013via,Cardona:2016ems,Zhang:2017aqn,Follin:2017ljs}. On the other hand, great efforts have been made to reconcile the two measurements by extending the base $\Lambda$CDM cosmology~\cite{Li:2013dha,Salvatelli:2013wra,Battye:2013xqa,Costa:2013sva,Zhang:2014ifa,Zhang:2014dxk,Battye:2014qga,
DiValentino:2016hlg,Qing-Guo:2016ykt,Bernal:2016gxb,DiValentino:2016ucb,Guo:2017hea,
Feng:2017nss,Zhao:2017urm,Guo:2017qjt,Benetti:2017gvm,DiValentino:2017zyq,DiValentino:2017iww,Yang:2017ccc,vanPutten:2017bqf,Feng:2017mfs,Zhao:2017jma, Feng:2017usu,Poulin:2018zxs,Guo:2018gyo,Yang:2018euj,Camarena:2018nbr,Choudhury:2018adz,DEramo:2018vss,Benetti:2017juy,Colgain:2018wgk,Choudhury:2018sbz,
Banihashemi:2018oxo,Carneiro:2018xwq}. For example, Refs.~\cite{Li:2013dha,Qing-Guo:2016ykt,Camarena:2018nbr} point out that a dynamical dark energy with $w<-1$ at low redshifts prefers a high value of $H_{0}$. Refs.~\cite{Battye:2013xqa,Zhang:2014ifa,Battye:2014qga,Zhang:2014dxk,Bernal:2016gxb,DiValentino:2016ucb,Guo:2017hea,Feng:2017nss,
Zhao:2017urm,Guo:2017qjt,Benetti:2017gvm,Feng:2017mfs,Zhao:2017jma,Benetti:2017juy,Choudhury:2018sbz,Carneiro:2018xwq} show that considering extra relativistic degrees of freedom $N_{\rm eff}$ in the $\Lambda$CDM model favors a high value of $H_{0}$ when $N_{\rm eff}>3.046$. In addition, considering a coupling between dark energy and dark matter also can affect the constraint results of $H_{0}$~\cite{Salvatelli:2013wra,Costa:2013sva,DiValentino:2017iww,Yang:2017ccc,Feng:2017usu,Guo:2018gyo,Yang:2018euj}. The impacts of these extra parameters on the fit value of $H_{0}$ can help relieve the $H_{0}$ tension. Obviously, considering all these extra parameters can definitely pull $H_{0}$ towards a higher value~\cite{DiValentino:2016hlg,DiValentino:2017zyq,Choudhury:2018adz}, relieving the $H_{0}$ tension to a great extent, but this does not mean that such an extension is favored by current observations. We thus wish to know if there is an extended model that can both relieve the $H_0$ tension and can be also favored by the current observations.


In this paper, we investigate several possibilities to reconcile the Planck data and the R16 result by considering extra parameters based on the $\Lambda$CDM model. We call these models extended $\Lambda$CDM (e$\Lambda$CDM) models. For these e$\Lambda$CDM models, they have at least one more parameter than the base $\Lambda$CDM model. By making comparison of constraint results of these e$\Lambda$CDM models, we wish to see if there exists a model that not only can reconcile the Planck data and the local measurement of $H_{0}$, but also is favored by current observations.
Of course, we do not mean that there are only these e$\Lambda$CDM models that can address the $H_0$ tension issue, and actually there are still a lot of other cosmological models that have been proposed to help mitigate the $H_{0}$ tension (see, e.g., Refs.~\cite{Bringmann:2018jpr,DiValentino:2017rcr,Lesgourgues:2015wza,Poulin:2016nat,Brust:2017nmv,Buen-Abad:2017gxg,Ko:2016uft,Ko:2016fcd,Ko:2017uyb,Poulin:2018cxd,Khosravi:2017hfi,Nunes:2018xbm,Lu:2016hsd,Alfaro:2018fmg}). But in this paper we only focus on these typical e$\Lambda$CDM models.

The structure of this paper is arranged as follows. In Sec.~\ref{sec:2}, various e$\Lambda$CDM models are briefly described. In Sec.~\ref{sec:3}, we introduce the observational data and the analysis method used in this paper. In Sec.~\ref{sec:4}, we give constraint results of these e$\Lambda$CDM models, and discuss if there is an extension to the $\Lambda$CDM model that is reasonable to relieve the tension between the Planck data and the local measurement of $H_{0}$. Conclusion of this work is drawn in Sec.~\ref{sec:5}.

\section{Extensions to the base $\Lambda$CDM cosmology}\label{sec:2}

We consider several extensions to the $\Lambda$CDM cosmology in order to relieve the $H_{0}$ tension. In this work, a spatially flat universe is considered, and thus the Friedmann equation is given by
\begin{equation}\label{2.1}
  H^{2}=\frac{8\pi G}{3} [\rho_{\rm r0}(1+z)^{4}+\rho_{\rm m0}(1+z)^{3}+\rho_{\rm de}(z)],
\end{equation}
where $\rho_{\rm r0}$ and $\rho_{\rm m0}$ are the current radiation density and matter density, respectively. The energy density of dark energy can be written as
\begin{equation}\label{2.2}
  \rho_{\rm de}(z)=\rho_{\rm de0} \exp \left\{3 \int^{z}_{0} \frac{dz^{\prime}}{(1+z^{\prime})} [1+w(z^{\prime})]\right\},
\end{equation}
where $\rho_{\rm de0}$ is the current dark energy density and $w$ is the equation-of-state parameter of dark energy.
In the $\Lambda$CDM model, the vacuum energy serves as dark energy which has $w=-1$. Six base parameters of this model include the energy densities of baryon $\omega_{\rm b}$ and cold dark matter $\omega_{\rm c}$, the acoustic angular scale $\theta_\ast$, the reionization optical depth $\tau$, and the amplitude $A_{\rm s}$ and the spectral index $n_{\rm s}$ of primordial scalar fluctuations.

For e$\Lambda$CDM cosmologies, we first consider three dynamical dark energy models, which are the $w$CDM model with a constant $w$, the holographic dark energy (HDE) model~\cite{Li:2004rb,Huang:2004mx,Zhang:2014ija} with $w(z)=-1/3-(2/3c)\sqrt{\Omega_{\rm de}(z)}$, where $c$ is a dimensionless parameter and the function $\Omega_{\rm de}(z)$ is determined by a differential equation (see Eq.~(18) in Ref.~\cite{Li:2004rb}), and the Chevallier-Polarski-Linder (CPL) model~\cite{Chevallier:2000qy,Linder:2002et} with $w(a) = w_{0}+w_{a}(1-a)$. The first two models are one-parameter extensions ($w$ for the $w$CDM model and $c$ for the HDE model), and the CPL model is a two-parameter extension ($w_{0}$ and $w_{a}$) to the $\Lambda$CDM model.

Then, we consider a coupling between dark energy and dark matter. We wish to extend the $\Lambda$CDM cosmology in this aspect, and thus we assume that vacuum energy interacts with cold dark matter. In this scenario, the vacuum energy density is no longer a pure background, but is a dynamical quantity. Such a model is called the $\Lambda (t)$CDM model or the I$\Lambda$CDM model. For more detailed introduction to the $\Lambda (t)$CDM model, see Refs.~\cite{Guo:2017hea,Feng:2017usu,Guo:2018gyo,Xiao:2018jyl}. In this paper, we take the energy transfer rate of $Q=\beta H\rho_{\rm c}$ as a typical example, where $\beta$ is a dimensionless coupling parameter, and in this model the parameter $\beta$ is the only extra parameter compared to the $\Lambda$CDM model. $\beta>0$ is defined as the case of cold dark matter decaying into the vacuum energy, and vice versa. To solve the large-scale instability problem of the interacting dark energy cosmology~\cite{Valiviita:2008iv}, we apply the extended parametrized post-Friedmann (PPF) approach for interacting dark energy cosmology~\cite{Guo:2017hea,Li:2014eha,Li:2014cee,Li:2015vla,Zhang:2017ize,Guo:2018gyo,Feng:2017usu,Feng:2018yew}. In Refs.~\cite{Guo:2017hea,Li:2015vla,Guo:2018gyo,Feng:2017usu}, it is shown that, under the extended PPF framework, we can explore the whole parameter space of the $\Lambda (t)$CDM model without any divergence of the perturbation of dark energy.

Next, we consider some other extensions to the $\Lambda$CDM cosmology. We consider the models with dark radiation (the effective number of relativistic species, $N_{\rm eff}$) and massive sterile neutrinos ($N_{\rm eff}$ and $m_{\nu, \rm sterile}^{\rm eff}$). Namely, we consider the $\Lambda$CDM+$N_{\rm eff}$ model and the $\Lambda$CDM+$N_{\rm eff}$+$m_{\nu, \rm sterile}^{\rm eff}$ model. Owing to a positive correlation between $N_{\rm eff}$ and $H_{0}$, the addition of the parameter $N_{\rm eff}$ in models can affect the constraint on $H_{0}$. It should be mentioned that, when sterile neutrinos are considered, we must have $N_{\rm eff}>3.046$, and in order to be distinct from the effects of cold or warm dark matter on the CMB, we assume $m_{\rm sterile}^{\rm thermal}<10$ eV, following the Planck collaboration \cite{Ade:2015xua}.


Finally, we would like to mention that some works have tried to relieve the $H_{0}$ tension by considering multi-parameter extensions. As shown in Ref.~\cite{Zhao:2017urm}, the $H_{0}$ tension can be relieved fairly well in the HDE model with sterile neutrinos. Refs.~\cite{Ade:2015xua,Battye:2013xqa,Battye:2014qga} indicate that the tension could be relieved in the $\Lambda$CDM model with both $\sum m_{\nu}$ and $N_{\rm eff}$. In this paper, we revisit constraints on these multi-parameter models. It should be pointed out that in this work we assume a normal hierarchy case for the neutrino mass in the $\Lambda$CDM+$\sum m_{\nu}$+$N_{\rm eff}$ model, for which the reason is that the current observations have evidently favored the normal hierarchy case of the neutrino mass over the inverted one, as shown in Refs.~~\cite{Guo:2018gyo,Huang:2015wrx,Wang:2016tsz,Yang:2017amu,Zhao:2017jma}.


\section{Data and method}\label{sec:3}



In this work, we use the Planck 2015 full-mission CMB temperature and polarization (TT, TE, EE) power spectra data, together with the Planck 2015 CMB lensing power spectrum data, presented in Ref.~\cite{Aghanim:2015xee}. In what follows, they are simply called ``CMB" data.

In addition, we employ the baryon acoustic oscillation (BAO) data and the ``Joint-Light Analysis" (JLA) sample of type Ia supernovae observation, to effectively break degeneracies among cosmological parameters. The BAO data include the measurements from the Date Release 12 of the SDSS-III Baryon Oscillation Spectroscopic Survey at $z_{\rm eff} = 0.32$ and $z_{\rm eff} = 0.57$~\cite{Gil-Marin:2015nqa}, the 6dF Galaxy Survey at $z_{\rm eff} = 0.106$~\cite{Beutler:2011hx}, and the Main Galaxy Sample of Data Release 7 of Sloan Digital Sky Survey at $z_{\rm eff} = 0.15$~\cite{Ross:2014qpa}. The JLA sample contains 740 type Ia supernovae data obtained from SNLS and SDSS as well as a few points of low redshift light-curve analysis~\cite{Betoule:2014frx}. The result of R16 ($H_{0}=73.00\pm1.75$ km s$^{-1}$ Mpc$^{-1}$)~\cite{Riess:2016jrr} is also combined with the CMB, BAO, JLA data, to obtain a high fit value of $H_{0}$.

It is worth mentioning that, recently, Riess et al. improved the result of R16 to $H_{0}=73.52\pm1.62$ km s$^{-1}$ Mpc$^{-1}$~\cite{Riess:2018byc} (hereafter R18) with a 2.3\% uncertainty. The tension is thus increased to $3.8\sigma$ between R18 and the 2015 Planck data (giving $H_{0}=66.93\pm0.62$ km s$^{-1}$ Mpc$^{-1}$ using the Planck TT, TE, EE+SIMLow data), but actually the uncertainty is only slightly reduced compared to R16. Hence, in our work, we still take the R16 measurement as a prior to combine other astrophysical data, which does not affect our discussion of the $H_{0}$ tension between the Planck data and the local measurement of the Hubble constant.

In our work, we employ the $\chi^{2}$ statistic method to perform the cosmological global fits. For each data set, we have $\chi^{2}_{\xi}=(\xi^{\rm obs}-\xi^{\rm th})^{2}/\sigma^{2}_{\xi}$, where $\xi^{\rm obs}$ and $\xi^{\rm th}$ are the experimentally measured value and the theoretically predicted value in cosmological models, respectively, and $\sigma^{2}_{\xi}$ is the standard deviation. Thus, in this paper, the total $\chi^{2}$ of the CMB+BAO+JLA+$H_{0}$ data can be written as
\begin{equation}\label{3.1}
  \chi^{2}= \chi^{2}_{\rm CMB}+\chi^{2}_{\rm BAO}+\chi^{2}_{\rm JLA}+\chi^{2}_{H_{0}}.
\end{equation}
In general, the $\chi^{2}$ comparison is sufficient and very popular for comparing different models with the same number of parameters. A smaller $\chi^{2}_{\rm min}$ means a better fit for a model.

However, for models with different number of parameters, a model with more parameters tends to lead to a smaller $\chi^{2}$. Under the circumstance, the $\chi^{2}$ comparison is unfair for comparing models. Thus, in this work, we also consider the Akaike information criterion (AIC) to compare different models. We have ${\rm AIC}=\chi^{2}_{\rm min}+2k$, where $k$ denotes the number of cosmological parameters. Actually, we only care about the relative value of the AIC between different models, i.e., $\Delta {\rm AIC}=\Delta \chi^{2}_{\rm min} +2\Delta k$. A model with a smaller value of $\rm {AIC}$ is more supported by current observations. In this work, the base $\Lambda$CDM model serves as a reference model. In general, we say that, compared to the reference model, a model with $0<\Delta \rm {AIC}<2$ is substantially supported, a model with $4<\Delta \rm {AIC}<7$ is considerably less supported, and a model with $\Delta \rm {AIC}>10$ is essentially not supported.

In this work, we modify the Boltzmann code {\tt CAMB}~\cite{Lewis:1999bs} to calculate the CMB power spectra for these e$\Lambda$CDM models, and also use the Markov-chain Monte Carlo package {\tt CosmoMC}~\cite{Lewis:2002ah} to explore the parameter spaces in these models (from which we can obtain the posterior distributions of parameters, as well as the best-fit points with $\chi^2_{\rm min}$, and 1$\sigma$ and 2$\sigma$ boundaries, etc). For details of the calculation methods, we refer the reader to Refs.~\cite{Lewis:1999bs,Lewis:2002ah}.

In the calculations, we assume flat priors for the cosmological parameters. In order not to affect the results of parameter estimation, we choose the prior ranges for the parameters to be much wider than the posteriors. For the 6 base parameters in the $\Lambda$CDM model, the prior ranges of them are chosen to be the same as those used by the Planck collaboration (see Table 1 in Ref.~\cite{Ade:2013zuv}). For the extra parameters in the e$\Lambda$CDM models, the prior ranges of them are: $[-3, 1]$ for $w$ in the $w$CDM model, $[0, 3]$ for $c$ in the HDE model, $[-3, 1]$ for $w_{0}$ and $[-4.5, 3.5]$ for $w_{a}$ in the CPL model, $[-0.3, 0.3]$ for $\beta$ in the $\Lambda (t)$CDM model, $[0, 6]$ for $N_{\rm eff}$ in the $\Lambda$CDM+$N_{\rm eff}$ model, $[0, 6]$ for $N_{\rm eff}$ and $[0.06, 3]$ eV for $\sum m_{\nu}$ in the $\Lambda$CDM+$\sum m_{\nu}$+$N_{\rm eff}$ model, and $[3.046, 7]$ for $N_{\rm eff}$ and $[0, 10]$ eV for $m_{\rm sterile}^{\rm thermal}$ in the $\Lambda$CDM (HDE)+$N_{\rm eff}$+$m_{\nu, \rm sterile}^{\rm eff}$ model.


\section{Results and discussion}\label{sec:4}

\begin{table*}[!htp]
\caption{Fit results of cosmological parameters in the $\Lambda$CDM model and the models with one more parameter than the $\Lambda$CDM model using the CMB+BAO+JLA+$H_{0}$ data. Here, the parameter $\alpha$ denotes $w$ in the $w$CDM model, $c$ in the HDE model, $\beta$ in the $\Lambda(t)$CDM model, and $N_{\rm eff}$ in the $\Lambda$CDM+$N_{\rm eff}$ model.}
\centering
\renewcommand{\arraystretch}{1.5}
\scalebox{0.9}[0.9]{%
\begin{tabular}{|c|c c c c c|}
\hline
 Model&$\Lambda$CDM &$w$CDM &HDE &$\Lambda (t)$CDM &$\Lambda$CDM+$N_{\rm eff}$  \\
\hline

$\Omega_{\rm b}h^2$      &$0 .02236\pm0.00014$
&$0 .02227\pm0.00015$
        &$0 .02243\pm0.00015$
        &$0 .02226\pm0.00016$
        &$0 .02249\pm0.00017$
 \\

$\Omega_{\rm c}h^2$     &$0 .1180\pm0.0010$
&$0 .1191\pm0.0012$
        &$0 .1169\pm0.0012$
        &$0 .1113^{+0.0040}_{-0.0041}$
        &$0 .1213\pm0.0027$
\\

$100\theta_{\emph{\rm MC}}$      &$1 .04102\pm0.00030$
&$1 .04088\pm0.00030$
        &$1 .04114\pm0.00030$
        &$1.04087\pm 0.00030$
        &$1 .04066\pm0.00039$
\\

$\tau$        &$0 .071\pm0.012$
&$0 .062\pm0.013$
        &$0 .089\pm0.014$
        &$0 .068\pm0.013$
        &$0 .071\pm0.012$
\\

${\textrm{ln}}(10^{10}A_{\rm s})$    &$3 .072\pm0.023$
&$3 .056^{+0.024}_{-0.025}$
        &$3 .106\pm0.025$
        &$3 .069\pm0.023$
        &$3 .080\pm0.023$
\\

$n_{\rm s}$   &$0 .9688\pm0.0039$
        &$0 .9658\pm0.0043$
        &$0 .9718\pm0.0044$
        &$0 .9653\pm0.0048$
        &$0 .9751\pm0.0063$
\\

$\alpha$         &$-$
                 &$-1.058\pm0.038$
                 &$0 .605^{+0.028}_{-0.031}$
                 &$0 .071^{+0.045}_{-0.044}$
                 &$3 .250\pm0.150$
                     \\

\hline

$\sigma_{8}$            &$0 .817\pm0.009$
         &$0 .830\pm0.012$
        &$0 .826\pm0.012$
        &$0 .844\pm0.019$
        &$0 .826\pm0.011$
        \\

$H_0$ [km/s/Mpc]                &$68 .09\pm0.45$
                                              &$69 .34\pm0.93$
                                              &$69 .67^{+0.95}_{-0.94}$
                                              &$69 .36\pm0.82$
                                              &$69 .25\pm0.99$
                                              \\

$H_0$ tension               &$2.72\sigma$
                                              &$1.85\sigma$
                                              &$1.67\sigma$
                                              &$1.88\sigma$
                                              &$1.87\sigma$
                                              \\

\hline
$\chi^2_{\rm min}$     &$13665.722$
                       &$13664.486$
                       &$13683.562$
                       &$13664.782$
                       &$13663.480$
                       \\

$\Delta {\rm AIC}$     &$0$
                       &$0.764$
                       &$19.840$
                       &$1.060$
                       &$-0.242$
                       \\
\hline
\end{tabular}}
\label{tab:l}
\end{table*}

\begin{table*}[!htp]
\caption{Fit results of cosmological parameters in the models with at least two more parameters than the $\Lambda$CDM model using the CMB+BAO+JLA+$H_{0}$ data.}
\centering
\renewcommand{\arraystretch}{1.5}
\scalebox{0.9}[0.9]{%
\begin{tabular}{|c|c c c c|}
\hline
 Model&CPL &$\Lambda$CDM+$\sum m_{\nu}$+$N_{\rm eff}$ &$\Lambda$CDM+$N_{\rm eff}$+$m_{\nu, \rm sterile}^{\rm eff}$&HDE+$N_{\rm eff}$+$m_{\nu, \rm sterile}^{\rm eff}$ \\
\hline

$\Omega_{\rm b}h^2$      &$0 .02224\pm0.00015$
&$0 .02254\pm0.00018$
        &$0 .02255^{+0.00017}_{-0.00019}$
        &$0 .02268^{+0.00020}_{-0.00022}$
 \\

$\Omega_{\rm c}h^2$     &$0 .1195\pm0.0013$
&$0 .1216^{+0.0027}_{-0.0028}$
        &$0 .1209\pm 0.0030$
        &$0 .1209^{+0.0035}_{-0.0028}$
\\

$100\theta_{\emph{\rm MC}}$      &$1 .04084\pm0.00031$
&$1 .04060\pm0.00040$
        &$1 .04064^{+0.00043}_{-0.00039}$
        &$1 .04067^{+0.00041}_{-0.00040}$
\\

$\tau$        &$0 .058^{+0.015}_{-0.014}$
&$0 .078^{+0.014}_{-0.015}$
        &$0 .079^{+0.014}_{-0.015}$
        &$0 .098\pm0.014$
\\

${\textrm{ln}}(10^{10}A_{\rm s})$    &$3 .049\pm0.027$
&$3 .094^{+0.027}_{-0.030}$
        &$3 .096^{+0.027}_{-0.029}$
        &$3 .134\pm0.029$
\\

$n_{\rm s}$   &$0 .9648\pm0.0045$
        &$0 .9775\pm0.0068$
        &$0 .9771^{+0.0069}_{-0.0078}$
        &$0 .9833^{+0.0085}_{-0.0088}$
\\

$w/w_{0}/c$         &$-1.000\pm0.100$
                    &$-$
                    &$-$
                    &$0 .627^{+0.035}_{-0.041}$
                     \\

$w_{a}$          &$-0.240^{+0.410}_{-0.340}$
                 &$-$
                 &$-$
                 &$-$
                     \\

$\sum m_{\nu}$ [eV]         &$-$
                 &$<0.22 $
                 &$-$
                 &$-$
                     \\

$N_{\rm eff}$         &$-$
                 &$3 .290\pm0.160$
                 &$<0.357$
                 &$<0.366$
                     \\

$m_{\nu, \rm sterile}^{\rm eff}$ [eV]           &$-$
                                            &$-$
                                            &$<0.359$
                                      &$<0.245$
                                      \\

\hline

$\sigma_{8}$            &$0 .830\pm0.012$
&$0 .820\pm0.012$
        &$0 .819^{+0.019}_{-0.013}$
        &$0 .828^{+0.017}_{-0.013}$
        \\

$H_0$ [km/s/Mpc]                &$69 .19^{+0.97}_{-0.96}$
                                              &$69 .20\pm1.00$
                                              &$69 .06^{+0.82}_{-1.17}$
                                              &$70.70\pm1.10$
                                              \\

$H_0$ tension
                                              &$1.90 \sigma$
                                              &$1.89 \sigma$
                                              &$1.88 \sigma$
                                              &$1.11 \sigma$
                                              \\

\hline
$\chi^2_{\rm min}$
                       &$13663.216$
                       &$13665.614$
                       &$13663.428$
                       &$13681.998$
                       \\

$\Delta {\rm AIC}$
                       &$1.494$
                       &$3.892$
                       &$1.706$
                       &$22.276$
                       \\
\hline
\end{tabular}}
\label{tab:2}
\end{table*}

\begin{figure*}[ht!]
\begin{center}
\includegraphics[width=7.0cm]{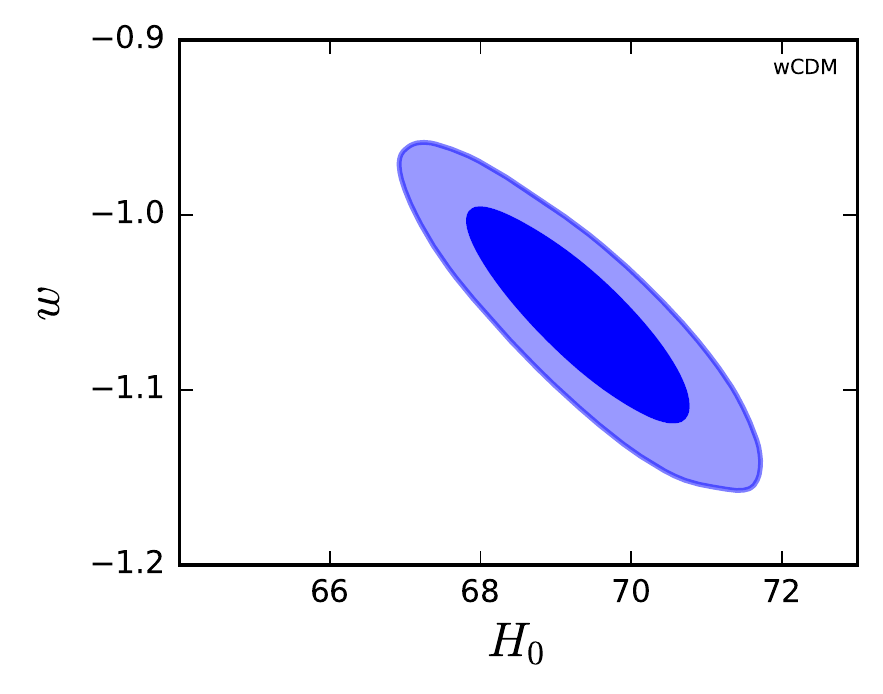}
\includegraphics[width=7.0cm]{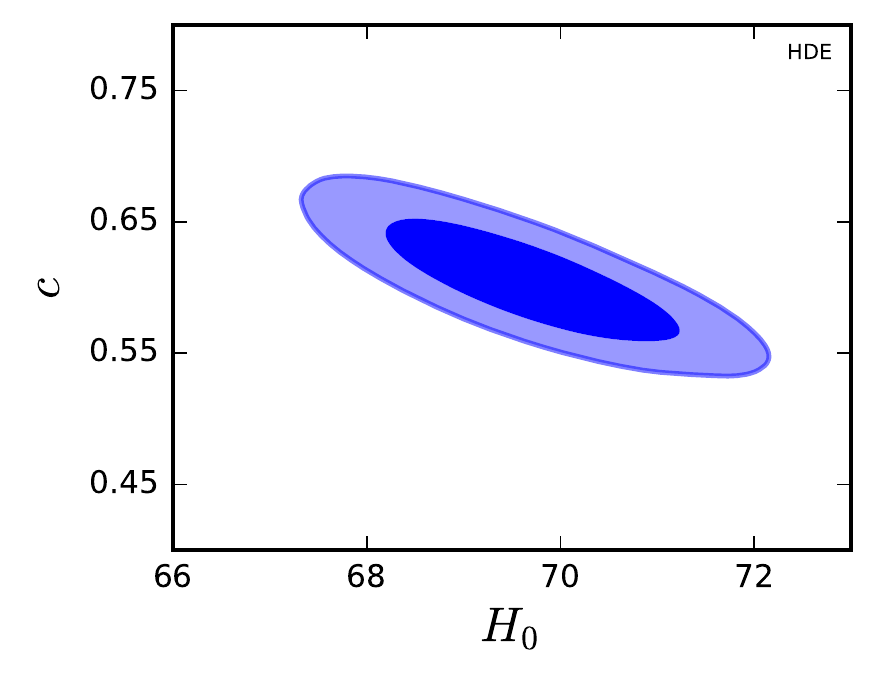}
\includegraphics[width=7.0cm]{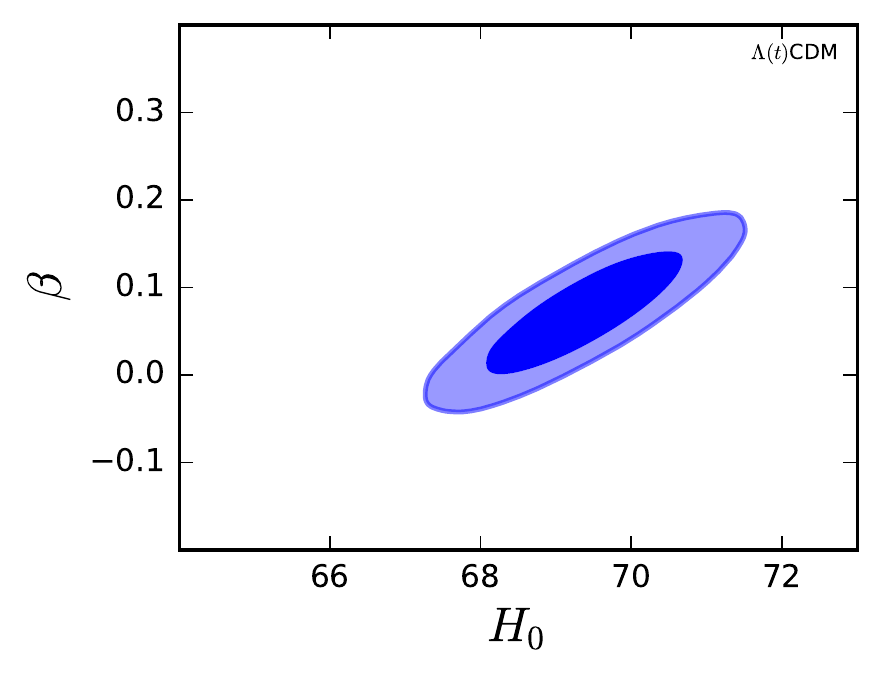}
\includegraphics[width=7.0cm]{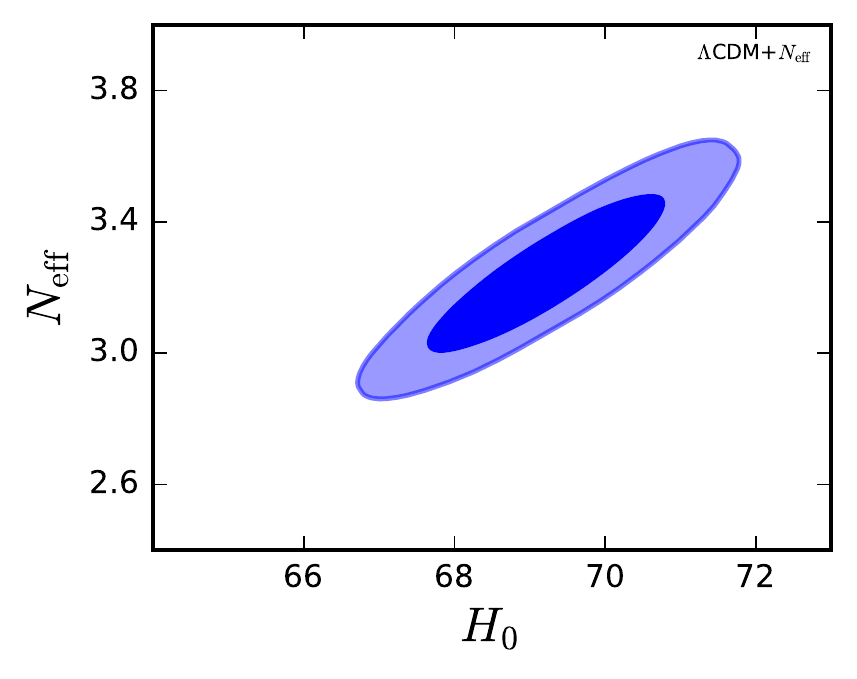}
\end{center}
\caption{Two-dimensional marginalized contours in the $w$--$H_{0}$ plane for the $w$CDM model, in the $c$--$H_{0}$ plane for the HDE model, in the $\beta$--$H_{0}$ plane for the $\Lambda (t)$CDM model, and in the $N_{\rm eff}$--$H_{0}$ plane for the $\Lambda$CDM+$N_{\rm eff}$ model, by using the CMB+BAO+JLA+$H_{0}$ data.}
\label{fig1}
\end{figure*}

We first constrain the base $\Lambda$CDM model using the CMB+BAO+JLA data and obtain $H_{0}=67.78^{+0.46}_{-0.45}$ km s$^{-1}$ Mpc$^{-1}$, which is $2.89 \sigma$ lower than R16.\footnote{The result of $H_{0}=67.78^{+0.46}_{-0.45}$ km s$^{-1}$ Mpc$^{-1}$ is more than $1\sigma$ higher than $H_{0}=66.93\pm 0.62$ km s$^{-1}$ Mpc$^{-1}$ \cite{Aghanim:2016yuo} (quoted by R16 \cite{Riess:2016jrr} to claim a 3.3$\sigma$ tension) derived from the Planck TT,TE,EE+SIMlow data, where SIMLow denotes the simulated CMB polarization data at low multipoles. The main reasons for this difference are: (i) Compared to the SIMLow data, the lowP data we employ in this paper prefer a higher value of $H_{0}$, as shown in Ref.~\cite{Aghanim:2016yuo}. (ii) The addition of the lensing data leads to a decrease in matter density, with an accompanying reduction in $\Omega_{\rm m}$. To keep $\theta_{*}$ (approximately $\propto \Omega_{\rm m} h^3$) fixed, an increase in $H_{0}$ is a must. In fact, our result of $H_{0}$ is consistent with those given by the Planck collaboration \cite{Ade:2015xua}. For example, the Planck collaboration \cite{Ade:2015xua} reported $H_{0}=67.27\pm 0.66$ km s$^{-1}$ Mpc$^{-1}$ with Planck TT,TE,EE+lowP and $H_{0}=67.74\pm 0.46$ km s$^{-1}$ Mpc$^{-1}$ with Planck TT,TE,EE+lowP+lensing+BAO+JLA+$H_{0}$ (here the local measurement of $H_{0}=70.6\pm3.3$ km s$^{-1}$ Mpc$^{-1}$ is used).} To obtain a higher $H_{0}$, we combine R16 with the CMB+BAO+JLA data, i.e., we also use the CMB+BAO+JLA+$H_{0}$ data. In this case, the result of $H_{0}=68.09 \pm 0.45$ km s$^{-1}$ Mpc$^{-1}$ is larger than the one from the CMB+BAO+JLA data, but is still $2.72\sigma$ lower than R16. In other words, there is still about $3\sigma$ tension between R16 and the global fit of $H_{0}$.

In Tables~\ref{tab:l} and~\ref{tab:2}, we show the constraint results of cosmological parameters and the tension between the best-fit $H_{0}$ and R16 in various e$\Lambda$CDM models. These models include the $w$CDM model, the HDE model, the $\Lambda (t)$CDM model, the $\Lambda$CDM+$N_{\rm eff}$ model, the CPL model, the $\Lambda$CDM+$\sum m_{\nu}$+$N_{\rm eff}$ model, the $\Lambda$CDM+$N_{\rm eff}$+$m_{\nu, \rm sterile}^{\rm eff}$ model, and the HDE+$N_{\rm eff}$+$m_{\nu, \rm sterile}^{\rm eff}$ model. In Table~\ref{tab:l}, model parameter $\alpha$ denotes $w$, $c$, $\beta$, and $N_{\rm eff}$ in different models. The $\chi^{2}_{\rm min}$ values and the $\Delta \rm {AIC}$ values are also shown in the two tables.

For the models with one more parameter than the $\Lambda$CDM model (see Table~\ref{tab:l}), we obtain $H_{0}=69.34\pm 0.93$ km s$^{-1}$ Mpc$^{-1}$ for the $w$CDM model, $H_{0}=69.67^{+0.95}_{-0.94}$ km s$^{-1}$ Mpc$^{-1}$ for the HDE model, $H_{0}=69.36\pm0.82$ km s$^{-1}$ Mpc$^{-1}$ for the $\Lambda (t)$CDM model, and $H_{0}=69.25\pm 0.99$ km s$^{-1}$ Mpc$^{-1}$ for the $\Lambda$CDM+$N_{\rm eff}$ model. Correspondingly, the $H_{0}$ tension between them and R16 are reduced to $1.85\sigma$ for the $w$CDM model, $1.67\sigma$ for the HDE model, $1.88\sigma$ for the $\Lambda (t)$CDM model, and $1.87\sigma$ for the $\Lambda$CDM+$N_{\rm eff}$ model, indicating that single-parameter extensions to the $\Lambda$CDM model can relieve the $H_{0}$ tension to some extent. Among these single-parameter extended models, the HDE model is the most effective one to relieve the $H_{0}$ tension.

However, compared with the base $\Lambda$CDM model, we find that the HDE model has $\Delta \chi^{2}_{\rm min}=17.840$ and $\Delta \rm {AIC}=19.840$, indicating that the HDE model is excluded by the current observations from a statistical point of view. For the $w$CDM model, the $\Lambda (t)$CDM model, and the $\Lambda$CDM+$N_{\rm eff}$ model, they all are favored by the current observations. We find that among these models the $\Lambda$CDM+$N_{\rm eff}$ model is most consistent with the current observational data, which has $\Delta \chi^{2}_{\rm min}=-2.242$ and $\Delta \rm {AIC}=-0.242$. This model can also effectively relieve the $H_{0}$ tension to be at less than (but still around) $2\sigma$ level.

Next, we give constraint results of multi-parameter extensions to the $\Lambda$CDM model. As can be seen from Table~\ref{tab:2}, we obtain $H_{0}=69.19^{+0.97}_{-0.96}$ km s$^{-1}$ Mpc$^{-1}$ for the CPL model, $H_{0}=69.20\pm1.00$ km s$^{-1}$ Mpc$^{-1}$ for the $\Lambda$CDM+$\sum m_{\nu}$+$N_{\rm eff}$ model, $H_{0}=69.06^{+0.82}_{-1.17}$ km s$^{-1}$ Mpc$^{-1}$ for the $\Lambda$CDM+$N_{\rm eff}$+$m_{\nu, \rm sterile}^{\rm eff}$ model, and $H_{0}=70.70\pm 1.10$ km s$^{-1}$ Mpc$^{-1}$ for the HDE+$N_{\rm eff}$+$m_{\nu, \rm sterile}^{\rm eff}$ model, indicating that the tensions with the R16 are at the $1.90\sigma$ level, the $1.89\sigma$ level, the $1.88\sigma$ level, and the $1.11\sigma$ level, respectively. We find that the HDE+$N_{\rm eff}$+$m_{\nu, \rm sterile}^{\rm eff}$ model can lead to a much smaller tension than the other three cases. However, it has $\Delta \chi^{2}_{\rm min}=16.276$ and $\Delta \rm {AIC}=22.276$, showing that the HDE+$N_{\rm eff}$+$m_{\nu, \rm sterile}^{\rm eff}$ model is actually excluded by the current observations. For the other three cases, we find that the $H_{0}$ tensions are about $\sim 1.9\sigma$, and according to their $\chi^2$ and AIC values these models are only slightly favored by the current observations.

Actually, from the constraint results of $H_{0}$ in these e$\Lambda$CDM models, we find that they all can relieve the $H_{0}$ tension to $2\sigma$ or less. As shown in Fig.~\ref{fig1}, considering these extra parameters in the $\Lambda$CDM model can affect the constraints on the Hubble constant $H_{0}$ because of having strong correlations between them, i.e., (i) $w$ and $c$ are anti-correlated with $H_{0}$, (ii) $\beta$ and $N_{\rm eff}$ are positively correlated with $H_{0}$. Synthetically speaking, from the statistical point of view, among these models, the case of considering $N_{\rm eff}$ in the $\Lambda$CDM is the best one to relieve the $H_{0}$ tension, which has $\Delta \chi^{2}_{\rm min}=-2.242$ and $\Delta \rm {AIC}=-0.242$. In the $\Lambda$CDM+$N_{\rm eff}$ model, a higher $H_{0}$ can be obtained when $N_{\rm eff}>3.046$. This is because a higher $N_{\rm eff}$ leads to a smaller sound horizon ($r_{\ast}$) at recombination. To keep the acoustic scale ($\theta_{\ast}$) fixed at the observed value, $H_{0}$ must rise ($\theta_{\ast}=r_{\ast}/D_{\rm A}$) to obtain a smaller angular diameter distance $D_{\rm A}$.

Although among these extended models the $\Lambda$CDM+$N_{\rm eff}$ model is the most preferred one and it can reduce the $H_0$ tension to be at the 1.87$\sigma$ tension, actually such a model cannot truly resolve the tension. This conclusion is drawn based on the following two facts. (i) The result is obtained by using the data combination of CMB+BAO+JLA+$H_{0}$ (R16). As discussed in the beginning of this section, the reason we use this data combination is to obtain a higher $H_0$. If we remove the $H_0$ prior of R16 in the data combination, i.e., we use the data combination of CMB+BAO+JLA, we will derive the results of $N_{\rm eff}=3.010^{+0.17}_{-0.18}$ and $H_{0}=67.50^{+1.20}_{-1.10}$ km s$^{-1}$ Mpc$^{-1}$, from which we can see that now the $H_0$ tension is at the 2.66$\sigma$ level. (ii) Increasing $N_{\rm eff}$ can indeed lead to a higher Hubble constant, but can also lead to a higher value of the fluctuation amplitude $\sigma_8$. In the $\Lambda$CDM model, we obtain $\sigma_8=0.817\pm0.009$, and in the $\Lambda$CDM+$N_{\rm eff}$ model, we obtain $\sigma_8=0.826\pm0.011$ (see Table~\ref{tab:l}). Hence, although a higher $N_{\rm eff}$ brings $H_0$ into better consistency with direct measurements, it also increases $\sigma_8$, aggravating the tension between the CMB measurements and astrophysical measurements of $\sigma_8$ discussed in Ref.~\cite{Ade:2015xua}. See also Fig. 31 in Ref.~\cite{Ade:2015xua} for relevant discussion. Similar analysis can also be found in Refs.~\cite{DiValentino:2016ucb,Battye:2013xqa,Zhao:2017urm,Zhao:2017jma}. It should be mentioned that in Refs. \cite{Bringmann:2018jpr,DiValentino:2017rcr,Lesgourgues:2015wza,Poulin:2016nat,Brust:2017nmv,Buen-Abad:2017gxg,Ko:2016uft,Ko:2016fcd,Ko:2017uyb} it was claimed that the $\sigma_{8}$ and $H_0$ tensions can be relieved simultaneously by considering a coupling between dark matter and dark radiation and also by combining the distance-redshift observations with large-scale structure observations. In this work, such a model is not considered and large-scale structure observations are not employed either. Therefore, by testing the various plausible extended models, we find that actually none of them can convincingly resolve the tension with R16 measurement of $H_0$.

\section{Conclusion}\label{sec:5}

We wish to investigate whether there is a plausible extension to the base $\Lambda$CDM cosmology that can resolve the tension between the Planck data and the R16 measurement ($H_{0}=73.00\pm1.75$ km s$^{-1}$ Mpc$^{-1}$).
We consider several single-parameter extensions including the $w$CDM model, the HDE model, the $\Lambda (t)$CDM model, and the $\Lambda$CDM+$N_{\rm eff}$ model. In addition, we also consider several multi-parameter extensions, such as the CPL model, the $\Lambda$CDM+$\sum m_{\nu}$+$N_{\rm eff}$ model, the $\Lambda$CDM+$N_{\rm eff}$+$m_{\nu, \rm sterile}^{\rm eff}$ model, and the HDE+$N_{\rm eff}$+$m_{\nu, \rm sterile}^{\rm eff}$ model. We combine the Planck 2015 CMB data with the BAO data and the JLA data to make the analysis. We find that in the $\Lambda$CDM model there is about $3\sigma$ tension between the CMB+BAO+JLA data and R16. Hence, we further use the CMB+BAO+JLA+$H_0$(R16) data combination to obtain a higher value of $H_{0}$, but we find that about $3\sigma$ tension still exists in the $\Lambda$CDM cosmology.

In the above extended cosmological models, we find that the $H_{0}$ tension indeed can be reduced to be at less than $2\sigma$ level, among which the HDE+$N_{\rm eff}$+$m_{\nu, \rm sterile}^{\rm eff}$ model and the HDE model are the most effective ones to relieve the $H_{0}$ tension ($1.11\sigma$ and $1.67\sigma$). But, from the statistical point of view, they are actually excluded by the current observations since their $\Delta \rm {AIC}> 10$ ($\Delta \rm {AIC}=19.840$ for the HDE model and $\Delta \rm {AIC}=22.276$ for the HDE+$N_{\rm eff}$+$m_{\nu, \rm sterile}^{\rm eff}$ model). By comparing the values of $\chi^{2}_{\rm min}$ and $\rm {AIC}$ of all these extended models, we find that the $\Lambda$CDM+$N_{\rm eff}$ model is the best one among these extended models to reconcile the Planck data with the local measurement of the Hubble constant. This model can relieve the $H_{0}$ tension to be at the $1.87\sigma$ level and it has $\Delta \chi^{2}_{\rm min}=-2.242$ and $\Delta \rm {AIC}=-0.242$. But actually even this model cannot truly resolve the $H_0$ tension (without the R16 $H_0$ prior in the data combination, the $H_0$ tension will be at the 2.66$\sigma$ level, and considering $N_{\rm eff}$ will increase the $\sigma_8$ tension).


In conclusion, by a careful test, we find that none of the extended cosmological models that we have investigated in this work can convincingly resolve the tension of the Planck 2015 data with the R16 measurement of the Hubble constant.

\vskip 0.3cm
{\bf Note added } Recently, the Planck 2018 results were released \cite{Aghanim:2018eyx}, which is the final release of the Planck mission. This work was done before the release of the Planck 2018 results. Also, the Planck 2018 data have still not been released. Thus, in this work, we use the Planck 2015 data to make an analysis.

\acknowledgments

This work was supported by the National Natural Science Foundation of China (Grants Nos.~11875102, 11835009, 11522540, and No.~11690021) and the Top-Notch Young Talents Program of China.

\end{document}